\documentclass[twocolumn]{aastex701}

\usepackage{bm}
\usepackage{CJK}

\begin{document}

\title{NERV: Neural-network Enhanced Reconstruction of the UniVerse with Application to Baryon Acoustic Oscillations in the BOSS DR12 Galaxy Sample}
\author[orcid=0000-0003-2299-6235, sname='Zhu']{Shi-Hui Zang\begin{CJK*}{UTF8}{gbsn}(臧诗慧)\end{CJK*}}
\affiliation{Department of Physics, University of Wisconsin-Madison, Madison, Wisconsin 53706, USA}
\email{szang4@wisc.edu}

\author[orcid=0000-0002-8202-8642, sname='Zhu']{Hong-Ming Zhu \begin{CJK*}{UTF8}{gbsn}(朱弘明)\end{CJK*}}
\affiliation{National Astronomical Observatories, Chinese Academy of Sciences, 20A Datun Road, Beijing 100101, China}
\affiliation{School of Astronomy and Space Science, University of Chinese Academy of Sciences, Beijing 100101, China}
\email[show]{hmzhu@nao.cas.cn}

\author[orcid=0000-0001-6772-9814, sname='Mao']{Tian-Xiang Mao \begin{CJK*}{UTF8}{gbsn}(毛天翔)\end{CJK*}}
\affiliation{National Astronomical Observatories, Chinese Academy of Sciences, 20A Datun Road, Beijing 100101, China}
\email{mtianxiang@gmail.com}

\author[orcid=0000-0003-2155-9578, sname='Pen']{Ue-Li Pen \begin{CJK*}{UTF8}{bsmi}(彭威禮)\end{CJK*}}
\affiliation{Institute of Astronomy and Astrophysics, Academia Sinica, Astronomy-Mathematics Building, No. 1, Sec. 4, Roosevelt Road, Taipei 106319, Taiwan}
\affiliation{Canadian Institute for Theoretical Astrophysics, 60 St. George Street, Toronto, Ontario M5S 3H8, Canada}
\affiliation{Canadian Institute for Advanced Research, 661 University Avenue, Toronto, Ontario M5G 1M1, Canada}
\affiliation{Dunlap Institute for Astronomy and Astrophysics, University of Toronto, 50 St. George Street, Toronto, Ontario M5S 3H4, Canada}
\affiliation{Perimeter Institute for Theoretical Physics, 31 Caroline Street North, Waterloo, Ontario N2L 2Y5, Canada}
\email{pen@cita.utoronto.ca}

\date{\today}

\begin{abstract}

We present the first application of neural-network-based baryon acoustic oscillation (BAO) reconstruction to real galaxy survey data, restoring the acoustic signature damped by nonlinear structure growth.
{\texttt{NERV}} ({\bf N}eural-network {\bf E}nhanced {\bf R}econstruction of the Uni{\bf V}erse) augments standard reconstruction with a convolutional neural network trained on cubic $N$-body simulations, and explicitly accounts for realistic observational effects including the curved-sky geometry, the redshift-dependent selection function, and finite survey boundaries, by tessellating the survey volume into local patches.
We validate the method on the \textsc{MultiDark-Patchy} mock catalogs, recovering unbiased BAO dilation parameters.
Applied to the BOSS DR12 galaxy sample, NERV improves the precision of the BAO distance measurements significantly.
These results establish neural reconstruction as a practical component of BAO analyses for ongoing surveys such as DESI, with the potential to substantially tighten constraints on the cosmic expansion history and the nature of dark energy.

\end{abstract}

\setcounter{tocdepth}{2}


\section{Introduction}

The $\Lambda$CDM model, in which dark energy is described by a cosmological constant, has been established as the standard model of cosmology through extensive tests with multiple generations of cosmological surveys \citep{WMAP:2012nax, Planck:2018vyg, BOSS:2016wmc, eBOSS:2020yzd}.
Recently, however, the combination of DESI baryon acoustic oscillation (BAO) measurements with other cosmological probes, including the cosmic microwave background (CMB) and Type Ia supernovae, has revealed a preference for dynamical dark energy over a cosmological constant, first reported at the $2.5$--$3.9\sigma$ level with the DR1 data \citep{DESI:2024mwx, Giare:2024oil} and strengthened to $2.8$--$4.2\sigma$ with DR2 \citep{DESI:2025zgx, Chaudhary:2025vzy}, potentially pointing to new physics beyond the standard $\Lambda$CDM framework.

In these measurements, the BAO feature serves as a theoretically clean and robust standard ruler for tracing the cosmic expansion history \citep{2dFGRS:2005yhx, SDSS:2005xqv}.
The feature originates from sound waves propagating in the tightly coupled photon--baryon plasma before recombination, which imprint a characteristic sound horizon at the drag epoch, $r_d \approx 150\,\mathrm{Mpc}$, on the late-time clustering of matter \citep{Peebles:1970ag, Sunyaev:1970bma, Eisenstein:1997ik}.
Because this scale is calibrated by well-understood linear physics in the early universe and lies far above the scales affected by astrophysical processes, the measured acoustic scale is shifted by nonlinear evolution and galaxy bias only at the sub-percent level \citep{Eisenstein:2006nj, Padmanabhan:2009yr}.
Nonlinear gravitational evolution, however, progressively broadens the acoustic peak. 
Large-scale bulk flows displace tracers from their initial positions and damp the oscillation amplitude, substantially degrading the attainable statistical precision even though the acoustic scale itself remains essentially unbiased \citep{Eisenstein:2006nj, Crocce:2007dt}.
Restoring the sharpness of the BAO feature is therefore essential for extracting the full cosmological information content of modern galaxy surveys.

Conventionally, this is achieved through a technique known as \textit{standard reconstruction} \citep{Eisenstein:2006nk}, which estimates the large-scale displacement field from the observed galaxy density field within the Zel'dovich approximation and moves the tracers backwards, thereby restoring part of the linear BAO information.
Since its first application to the SDSS luminous red galaxy sample \citep{Padmanabhan:2012hf}, standard reconstruction has become a routine component of BAO analyses in modern galaxy surveys \citep{BOSS:2016wmc, BOSS:2016hvq, eBOSS:2020yzd, DESI:2024mwx, DESI:2025zgx}.
To push beyond this linear treatment, a variety of more sophisticated techniques have been proposed, including nonlinear reconstruction \citep{Zhu:2016sjc, Shi:2017gqs} and iterative initial-condition reconstruction \citep{Schmittfull:2017uhh, Hada:2018fde}, which aim to recover the displacement or initial density field with higher fidelity.
Nevertheless, the performance of these estimators remains limited by the complex nonlinear evolution of the density field and by galaxy bias \citep{Ota:2021caz, Ota:2022him}.
Moreover, while standard reconstruction has mature procedures for handling survey boundaries, selection functions, and spatially varying line of sight \citep{Burden:2015pfa, BOSS:2016sne}, these advanced methods have so far been validated primarily in idealized simulation volumes, and their application to realistic survey data remains an open challenge.

Machine learning provides an alternative route to capturing the nonlinear information.
Convolutional neural networks (CNNs) trained on simulations have been shown to recover the initial density field more accurately than standard reconstruction, learning the nonlinear mapping between the late-time and initial density fields directly from the data \citep{Mao:2020vdp, Shallue:2022mhf, Chen:2023uup}.
Rather than explicitly estimating and reversing a displacement field, the network exploits higher-order and small-scale information that is inaccessible to estimators built on the Zel'dovich approximation.
More recently, hybrid schemes that apply a CNN correction on top of standard reconstruction have demonstrated further gains in the recovery of linear modes and BAO information \citep{2025JCAP...09..039P, Bayer:2026zcr}.
To date, however, all of these approaches have been developed and validated in periodic simulation boxes. 
Applying them to observational data remains challenging owing to the curved-sky geometry, redshift-dependent selection functions, and irregular survey boundaries that have no counterpart in the training simulations.
Bridging this gap requires a reconstruction framework that retains the nonlinear information learned by the network while accommodating the geometry and selection effects of realistic galaxy surveys.

In this paper, we present such a framework: NERV (Neural-network Enhanced Reconstruction of the UniVerse), a hybrid approach that couples standard reconstruction with a CNN-based enhancement and handles the curved-sky geometry of real surveys by tessellating the survey volume into local patches.
Using controlled numerical experiments, we systematically evaluate the observational effects present in realistic data including the spatially varying line of sight, the redshift-dependent number density, and survey boundaries.
We also carefully quantify how mismatches between the training simulations and the target data, in redshift and in galaxy number density, propagate into the reconstructed field.
We train the network on mock galaxy catalogs constructed from the \textsc{AbacusSummit} simulations \citep{Maksimova:2021ynf} and validate the full pipeline on the \textsc{MultiDark-Patchy} mock catalogs \citep[][]{Kitaura:2015uqa}, recovering unbiased BAO dilation parameters.
Finally, we apply NERV to the BOSS DR12 galaxy sample.
To our knowledge, the first application of neural-network-based BAO reconstruction to real galaxy survey data.
In the lower-redshift bin, we obtain up to a $\sim25\%$ improvement in the precision of the BAO distance measurements relative to standard reconstruction, while the higher-redshift bin shows no improvement.
When extended to ongoing surveys such as DESI, such gains translate into a substantial increase in effective survey volume and will directly sharpen the BAO input to the dark energy constraints discussed above.

This paper is organized as follows.
Section~\ref{sec:method} describes the NERV framework.
Section~\ref{sec:robustness} presents controlled tests of the observational effects and training--data mismatches relevant to realistic surveys.
In Section~\ref{sec:results} we present the validation on mock catalogs and the BAO measurements from the BOSS DR12 galaxy sample.
Finally, Section~\ref{sec:discussion} discusses the implications of our results for future surveys.

\section{Method}
\label{sec:method}
In this section, we describe the details of NERV framework. 
NERV consists of two stages: a conventional standard-reconstruction step, followed by a CNN-based enhancement that takes as input both the original galaxy overdensity field and the standard-reconstruction output, plus a sky tessellation scheme that deals with curved sky geometry.
Standard reconstruction reverses the large-scale bulk flows, which are governed by well-understood quasi-linear dynamics but coherent over distances beyond the receptive field of a practical network, while the CNN supplies the spatially local corrections for the residual nonlinear evolution that the linear treatment cannot capture.
Similar hybrid strategies have proven effective in periodic simulation volumes \citep[e.g.][]{2025JCAP...09..039P, Bayer:2026zcr}.
NERV extends this design to realistic survey geometry by performing the CNN stage on a tessellation of the survey volume into local patches.
In the following, we first introduce the data sets used for training and inference, then describe the two reconstruction stages, and finally outline the power spectrum estimation and the BAO fitting methodology.

\begin{figure*}
    \centering
    \includegraphics[width=0.87\linewidth]{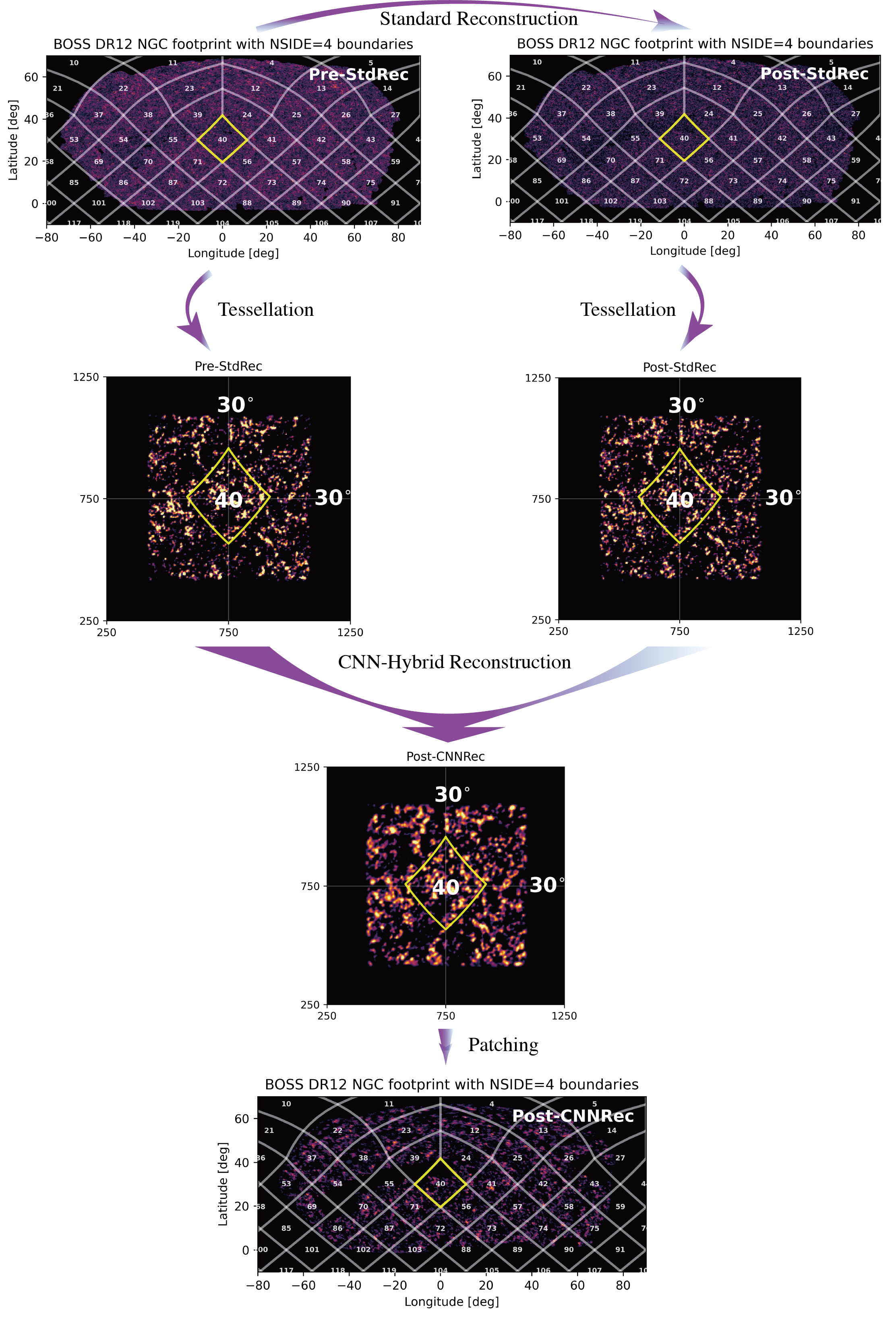}
    \caption{The flowchart of the NERV framework.
            In the first stage, the pre- and post-standard-reconstruction galaxy density fields are tessellated into multiple local patches.
            The CNN is then applied independently to each pair of local density fields to recover the corresponding reconstructed density field.
            Finally, the reconstructed patches are combined to form the full survey density field used for power-spectrum estimation.}
    \label{fig:CNNRec_process}
\end{figure*}

\subsection{Data Sets}
\label{sec:data}
\subsubsection{BOSS DR12 Galaxy Catalog}
In this work, we use the BOSS Data Release 12 (DR12) galaxy catalog as our observational data set
\citep{BOSS:2011sdu, BOSS:2012dmf, BOSS:2015ewx}.
We additionally use the \textsc{MultiDark-Patchy} mock catalogs to estimate the covariance matrices.
Both the BOSS data and the mock catalogs are divided into two redshift bins,
$(z_{\mathrm{min}},z_{\mathrm{max}})
=\{(0.2,0.5),(0.5,0.75)\}$,
with corresponding effective redshifts
$z_{\mathrm{eff}}=0.38$ and $0.61$, respectively.
We adopt the standard BOSS systematic weights $w_c = (w_{\mathrm{rf}} + w_{\mathrm{fc}} - 1)w_{\mathrm{sys}}$ throughout the analysis where $w_{\mathrm{rf}}$ is the redshift failure weight, $w_{\mathrm{fc}}$ is the fibre collision weight and $w_{\mathrm{sys}}$ is systematics weight.
A total of $N_{\mathrm{sim}}=500$ mock realizations are used to estimate the covariance matrices of the power-spectrum multipoles for both the pre- and post-reconstruction measurements.
Throughout the analysis, we adopt a flat $\Lambda$CDM fiducial cosmology with $\Omega_{\mathrm{m}}=0.3152$ and $h=0.6736$ to convert observed redshifts and angles into comoving Cartesian coordinates.
This fiducial model matches the base cosmology of the \textsc{AbacusSummit} simulations used for training, ensuring that no cosmology mismatch is introduced between the training and inference stages.

The galaxy overdensity field is calculated from the galaxy and random catalogs via
\begin{equation}
    \delta_g(\bm{x})
    =
    \frac{G(\bm{x})}{\alpha R(\bm{x})}
    - 1,
\end{equation}
where $G(\bm{x})$ and $R(\bm{x})$ denote the weighted galaxy and random number-density fields respectively.
The normalization factor is defined as
\begin{equation}
    \alpha = \frac{N_{\mathrm{gal}}}{N_{\mathrm{ran}}},
\end{equation}
where $N_{\mathrm{gal}}$ and $N_{\mathrm{ran}}$ are the total weighted numbers of galaxies and random particles.

\subsubsection{Training Set}

We use an additional set of mock galaxy catalogs constructed from the \textsc{AbacusSummit} simulations \citep[][]{Maksimova:2021ynf} to train the neural network.
Specifically, we employ the first 24 of 25 base-resolution simulations at $z=0.5$.
Dark-matter haloes are identified using the \textsc{CompaSO} halo finder \citep[][]{Hadzhiyska:2021zbd}, and galaxies are subsequently populated with \textsc{AbacusHOD} \citep[][]{Yuan:2021izi} to approximately reproduce the properties of the observed galaxy sample.

The resulting galaxy catalogs have a mean number density of
$\bar{n}=8.5\times10^{-4}\,h^{3}\mathrm{Mpc}^{-3}$.
Because the two redshift bins of the BOSS DR12 sample have different characteristic number densities, we train a separate network for each bin.
For the first redshift bin ($z_{\mathrm{eff}}=0.38$), we randomly downsample the training catalogs to
$\bar{n}=5.0\times10^{-4}\,h^{3}\mathrm{Mpc}^{-3}$,
approximately matching the maximum number density of this bin.
For the second redshift bin ($z_{\mathrm{eff}}=0.61$), we downsample instead to
$\bar{n}=2.0\times10^{-4}\,h^{3}\mathrm{Mpc}^{-3}$,
matching the lower characteristic number density of this higher-redshift bin.
Finally, the galaxy distributions are assigned to $512^3$ Cartesian grids using the Triangular-Shaped-Cloud (TSC) mass-assignment scheme, corresponding to a grid spacing of $2000/512\approx3.9\,h^{-1}\mathrm{Mpc}$.

\subsection{Standard Reconstruction}

In the first stage, we employ the public code \texttt{pyrecon}\footnote{\url{https://github.com/cosmodesi/pyrecon}} to implement standard reconstruction, which aims to reverse the large-scale linear displacements of galaxies, $\bm{\Psi}(\bm{x})$, based on the Zel'dovich approximation. 
The displacement field is obtained by solving
\begin{equation}
\nabla \cdot \bm{\Psi} + f \, \nabla \cdot \bigl( \bm{\Psi} \cdot \hat{\bm{r}} \bigr) \, \hat{\bm{r}} = -\frac{\delta_g}{b},
\label{eq:Zeldovich_approx}
\end{equation}
where $f$ denotes the linear growth rate, $\hat{\bm{r}}$ is the unit vector along the line of sight, $\delta_g$ is the observed galaxy overdensity field, and $b$ is the galaxy bias. In simulations with a fixed line of sight, this equation can be solved straightforwardly in Fourier space. 
However, for realistic survey geometries in the curved sky, where the line of sight varies between the survey volumes, we adopt the \textsc{IterativeFFT} approach \citep[][]{Burden:2015pfa}, which can better handle the varying lines of sight directions.
We use a smoothing scale of $15\ h^{-1}\mathrm{Mpc}$ for standard reconstruction, which is performed on a Cartesian grid with $1024^3$ cells in a box of side length $5000\ h^{-1}\mathrm{Mpc}$.
The galaxy bias $b$ and linear growth rate $f$ entering Equation~\ref{eq:Zeldovich_approx} are set to the values adopted in the BOSS DR12 Fourier-space BAO analysis \citep{BOSS:2016hvq}.
After solving for the displacement field, we follow the \textsc{RecSym} convention \citep[see, e.g.,][]{Chen:2024tfp} where both galaxies and randoms are shifted by the full redshift-space displacement, including the line-of-sight component, so that large-scale redshift-space distortions are retained in the reconstructed field.

\subsection{CNN Enhancement}
In the second stage, a convolutional neural network refines the standard-reconstruction output.
The network takes as input both the observed galaxy overdensity field $\delta_g$ and the post-reconstruction field $\delta_g^{\mathrm{rec}}$, and is trained to predict the linear density field from their combination.
Because the network is trained on periodic cubic boxes drawn from the \textsc{AbacusSummit} simulations (Section~\ref{sec:data}) with a fixed line of sight, it cannot be applied to a wide-field survey directly.
NERV therefore tessellates the survey volume into a set of local patches, within each of which a fixed line of sight is a good approximation, and applies the same trained network independently to every patch.
In the following, we describe the network architecture and training procedure, followed by the curved-sky tessellation used at inference.

\subsubsection{Network Architecture and Training}
\begin{figure*}
    \centering
    \includegraphics[width=0.9\linewidth]{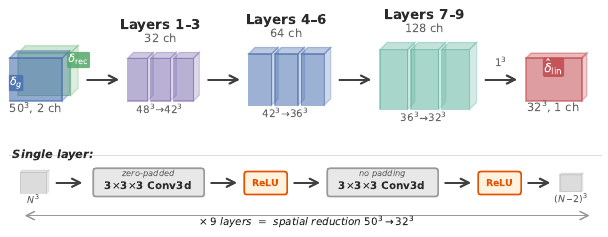}
    \caption{The neural network architecture. 
    The network takes as input a $50^3$-voxel subgrid with two channels, the galaxy overdensity field $\delta_g$ and the standard reconstruction result $\delta_{\rm rec}$ and outputs a $32^3$ voxel predicted linear density field $\hat{\delta}_{\rm lin}$.
    The model consists of 9 layers of paired zero-padded and valid $3\times3\times3$ convolutions with ReLU activation, each reducing the grid dimension by two voxels. Channels increase from 32 to 64 to 128 across three blocks before a $1^3$ projection. 
    For the inference, overlapping subgrids are tiled across the volume and averaged.}
    \label{fig:cnn_architecture}
\end{figure*}

We adopt the same neural network architecture used by \citet[][]{2025JCAP...09..039P}, which is illustrated in Figure~\ref{fig:cnn_architecture} .
The network consists of nine consecutive convolutional blocks, each containing two $3\times3\times3$ convolutional layers followed by ReLU activations.
The input is a cubic subgrid of size $N_{\mathrm{sub}}^3=50^3$ with two channels, corresponding to the galaxy overdensity field $\delta_g$ and the standard-reconstructed field $\delta_g^{\mathrm{rec}}$.
As the input propagates through the convolutional blocks, the spatial extent is progressively reduced, resulting in an output field of size $(N_{\mathrm{sub}}-18)^3=32^3$.

To train the network, we adopt the weighted Fourier-space loss function introduced by \citet{2025JCAP...09..039P}:
\begin{equation}
\mathcal{L}
=
\sum_{\mathbf{k}}
M(k)
\left|
\tilde{
f}\!\left(\delta_g,\delta_g^{\mathrm{rec}}\right)
(\mathbf{k})
-
\tilde{\delta}_{\ell}(\mathbf{k})
\right|^2 ,
\end{equation}
where $f$ denotes the neural network and $\delta_{\ell}$ is the target linear density field.
The weighting function $M(k)$ upweights modes in the range
$k\in[0.08,0.5]\,h\,\mathrm{Mpc}^{-1}$ by a factor of 10 and is set to unity otherwise.
This weighting enhances the relative contribution of the modes that carry the BAO signature and the quasi-linear information targeted by the enhancement, following Eqs.~(2.8) and (2.9) of \citet{2025JCAP...09..039P}.

\subsubsection{Curved-Sky Tessellation}
To apply the trained CNN to curved-sky survey data, we construct a set of local Cartesian grids covering small angular regions on the sky.
We first tessellate the survey footprint using the \textsc{HEALPix} scheme with \texttt{NSIDE}=4\footnote{We have verified that using \texttt{NSIDE}=3 or \texttt{NSIDE}=5 leads to nearly identical power-spectrum monopoles and quadrupoles, indicating that the results are insensitive to this choice.}, corresponding to 192 pixels with a characteristic angular scale of approximately $14.7^\circ$.
Each \textsc{HEALPix} pixel defines the center of a local patch.
To reduce boundary effects in the CNN inference, we include galaxies and randoms within $15.0^\circ$ of the patch center, such that the actual CNN input region extends beyond the corresponding \textsc{HEALPix} pixel.
A more detailed discussion of the boundary treatment is presented in Section~\ref{sec:boundary}.
For each local patch, the galaxy and random catalogs are mapped onto a $384^3$ Cartesian grid with a side length of $L=1500\,h^{-1}\mathrm{Mpc}$.
The box size is chosen to fully contain the objects in each local patch, while the resulting grid spacing exactly matches that of the training grids.
Within each patch, the trained network is applied to overlapping $50^3$ subgrids tiled across the grid, and the overlapping outputs are averaged to form the reconstructed density field of the patch.

\subsection{Power Spectrum Estimator}
\label{sec:pk}

We estimate the power-spectrum multipoles using the Yamamoto estimator
\citep[][]{Yamamoto:2005dz}, following the FFT-based implementations of
\citet[][]{Bianchi:2015oia} and \citet[][]{Scoccimarro:2015bla}.
This estimator accounts for the spatially varying line of sight in a wide-angle galaxy survey and is implemented using the public code \texttt{pypower}\footnote{\url{https://github.com/cosmodesi/pypower}}.

For the standard reconstruction, the power spectrum is estimated directly from the galaxy and random catalogs.
The NERV, however, produces a continuous density field on a set of local Cartesian grids rather than a discrete galaxy catalog.
To estimate its power spectrum, we interpret each grid cell as an effective particle whose weight is given by the reconstructed overdensity in that cell.
In the overlap regions between adjacent patches, each grid cell is taken exclusively from the patch whose \textsc{HEALPix} pixel contains it, so the extended patch boundaries serve only as input padding for the network and no volume is double counted.
The effective particles from all local patches are then combined and deposited onto a common full-survey Cartesian grid using the Nearest-Grid-Point (NGP) mass-assignment scheme, after which the power-spectrum multipoles are measured with the same Yamamoto estimator.

This procedure leaves the overall normalization of the reconstructed power spectrum arbitrary.
The normalization has no impact on the BAO measurement since it is absorbed by the amplitude parameters of the template and by the broadband terms marginalized over in the fitting process, leaving the inferred acoustic scale unaffected.

Following the convention adopted in the BOSS DR12 Fourier-space BAO analysis \citep{BOSS:2016hvq}, we measure the power-spectrum multipoles in linearly spaced bins of width $\Delta k=0.01\,h\,\mathrm{Mpc}^{-1}$ over $k\in[0.01,0.3]\,h\,\mathrm{Mpc}^{-1}$, yielding 29 bins per multipole.
We additionally apply the FKP weight \citep{Feldman:1993ky} to each grid cell according to the local redshift-dependent number density,
\begin{equation}
    w_{\mathrm{FKP}}(z)
    =
    \frac{1}{1+n(z)P_0},
\end{equation}
where $P_0=10 \ 000\,h^{-3}\mathrm{Mpc}^{3}$.

\subsection{Fitting the BAO signal}
\label{sec:baofit}
In Fourier space, the BAO signal manifests as a series of oscillations in the power spectrum with a period of $2\pi/r_d$ in wavenumber, where $r_d$ denotes the sound horizon at the baryon-drag epoch. Modeling these oscillations enables measurements of the combinations $D_A(z)/r_d$ in the transverse direction and $H(z)\,r_d$ along the line of sight, where $D_A(z)$ is the angular diameter distance and $H(z)$ is the Hubble parameter at redshift $z$.
In modern galaxy surveys, power spectrum multipoles are typically computed assuming a fiducial cosmology and subsequently fitted with a BAO template to extract two dilation parameters,
\begin{equation}
\alpha_\parallel = \frac{H^{\mathrm{fid}} \, r_d^{\mathrm{fid}}}{H(z) \, r_d},
\qquad
\alpha_\perp = \frac{D_A(z) \, r_d^{\mathrm{fid}}}{D_A^{\mathrm{fid}} \, r_d},
\end{equation}
where the superscript ``fid" denotes quantities evaluated in the fiducial cosmology. 
The uncertainties on $\alpha_\parallel$ and $\alpha_\perp$ therefore quantify the constraining power of the BAO measurement along the line-of-sight and transverse directions, respectively.

Note that the fiducial cosmology used throughout this work, for training, inference, and the redshift-to-distance conversion, is that of the \textsc{AbacusSummit} simulations, whereas the \textsc{MultiDark-Patchy} mocks were generated with a slightly different cosmology.
The expected dilation parameters for the mocks therefore deviate from unity: $\alpha_\parallel^{\mathrm{true}}=0.994$ and $\alpha_\perp^{\mathrm{true}}=0.992$ at $z_{\mathrm{eff}}=0.38$, and $\alpha_\parallel^{\mathrm{true}}=0.996$ and $\alpha_\perp^{\mathrm{true}}=0.993$ at $z_{\mathrm{eff}}=0.61$.

In this work, we follow the template by \citet{Chen:2024tfp}, as implemented in the public \texttt{desilike} package\footnote{\url{https://github.com/cosmodesi/desilike}}.
Within this model, the galaxy power spectrum is given by
\begin{equation}
P_{gg}(k,\mu) = \mathcal{B}(k,\mu)\,P_{\rm nw}(k) + \mathcal{C}(k,\mu)\,P_{\rm w}(k) + \mathcal{D}_\ell(k),
\end{equation}
where $P_{\rm w}(k)$ and $P_{\rm nw}(k)$ denote the linear wiggle and no-wiggle power spectra in the fiducial cosmology.
The second term $\mathcal{C}(k, \mu)P_{\rm w}(k)$ isolates the BAO signal, while the remaining terms account for the broadband contribution.
In particular, $\mathcal{B}(k,\mu)\,P_{\rm nw}(k)$ captures the smooth, non-oscillatory component of the linear power spectrum, and $\mathcal{D}_\ell(k)$ absorbs residual contributions arising from nonlinear evolution and higher-order effects that are not fully described by the first two terms.
Specifically, we have
\begin{equation}
    \mathcal{B}(k,\mu) =  (b + f\mu^2)^2 \left(1 + \frac{1}{2}k^2\mu^2\Sigma_s^2\right)^{-2},
\end{equation}
\begin{equation}
\mathcal{C}(k,\mu) =  (b + f\mu^2)^2\exp\!\left[-\frac{1}{2}k^2\bigl(\mu^2\Sigma_\parallel^2+(1-\mu^2)\Sigma_\perp^2\bigr)\right],
\end{equation}
where $(b + f\mu^2)$ is the linear Kaiser factor. The second factor in $\mathcal{B}(k,\mu)$ models Finger-of-God (FoG) damping along the line of sight, with $\Sigma_s$ treated as a free velocity-dispersion parameter. 
The parameters $\Sigma_\parallel$ and $\Sigma_\perp$ describe the BAO damping scales parallel and perpendicular to the line of sight, respectively, and are varied in the fit.
The residual term $\mathcal{D}_\ell(k)$ is modeled using a cubic spline basis,
\begin{equation}
\mathcal{D}_\ell(k) = \sum_{n=-1}^{{n_{\max}}} a_{\ell,n}{\,}W_3\left(\frac{k}{\Delta} - n\right),
\end{equation}
where $W_3$ is the piecewise cubic spline extension of the counts-in-cell interpolation kernel \citep[][]{1988csup.book.....H, 2010PhDT.........4J}, and the coefficients $a_{\ell,n}$ are marginalized over.
The node spacing $\Delta$ sets the smoothness of the broadband component, and $n_{\max}$ is the number of nodes required to span the fitted $k$ range.

Finally, we combine the BAO template with the covariance matrix $C_{ij}$ estimated from the \textsc{MultiDark-Patchy} mock catalogs to constrain the dilation parameters $\alpha_\parallel$ and $\alpha_\perp$.
Parameter inference is performed assuming a Gaussian likelihood,
$\mathcal{L}\propto\exp(-\chi^2/2)$, with
\begin{equation}
\chi^2 =
\frac{N_{\mathrm{sim}}-n_b-2}{N_{\mathrm{sim}}-1}
\sum_{i,j}
\left(P_i^{\mathrm{obs}}-P_i^{\mathrm{model}}\right)
C^{-1}_{ij}
\left(P_j^{\mathrm{obs}}-P_j^{\mathrm{model}}\right),
\end{equation}
where the indices $i$ and $j$ run over the elements of the power-spectrum data vector.
The prefactor is the Hartlap correction \citep[][]{Hartlap:2006kj}, which accounts for the bias in the inverse sample covariance arising from the finite number of mock realizations.
In our analysis, we use $N_{\mathrm{sim}}=500$ mocks and a data-vector length of $n_b=58$.

Before evaluating the likelihood, the theoretical power-spectrum model is convolved with the survey window function to account for mode coupling induced by the survey geometry, following \citet[][]{2017MNRAS.464.3121W} and \citet[][]{BOSS:2016hvq}.

\section{Robustness Tests}
\label{sec:robustness}

NERV is trained under idealized conditions: periodic boxes, a fixed line of sight, and a uniform galaxy sample.
A real survey is different in every one of these respects.
The line of sight varies across the sky, the galaxy number density changes with redshift, and the survey volume ends at boundaries.
In addition, the training simulations may differ from the observed sample, for example in redshift and in number density.

Before applying NERV to the BOSS DR12 data, we test the impact of each of these differences using controlled numerical experiments, in which the effects can be switched on one at a time.
Section~\ref{sec:obs_eff} covers the observational effects, and Section~\ref{sec:mod_dis} covers the training--data mismatches.
In all cases we find that NERV remains robust.

\subsection{Observational Effects}
\label{sec:obs_eff}

Each of the three observational effects changes the density field that the network receives as input, and may bias the reconstruction if left untreated.
Our tests start from the fiducial simulation \texttt{AbacusSummit\_base\_c000\_ph024}, which we divide into eight subboxes with a side length of $1000\,h^{-1}\mathrm{Mpc}$ each.
Each test uses one or more of these subboxes, so that a single effect can be isolated and examined before the method is applied to the BOSS DR12 data.

\subsubsection{Effect of the Varying Line of Sight}
In cubic simulations, the line of sight is usually fixed under the plane-parallel approximation. 
Combined with periodic boundary conditions, this preserves translational invariance, so the same neural network can be applied to different subvolumes by sliding a subgrid window across the simulation.

In real surveys, the line of sight varies across the sky because it is defined by the position of each galaxy relative to the observer. 
This breaks the exact translational invariance assumed in cubic simulations. 
However, within sufficiently small angular patches, this variation is limited, allowing NERV to be applied locally with an approximately fixed line of sight.
The relevant question is therefore how small the local patches need to be for the plane-parallel approximation to remain accurate, and whether the residual line-of-sight variation within each local patch induces a significant bias in the reconstructed density field.

\begin{figure*}[htbp]
    \centering
    \includegraphics[width=\linewidth]{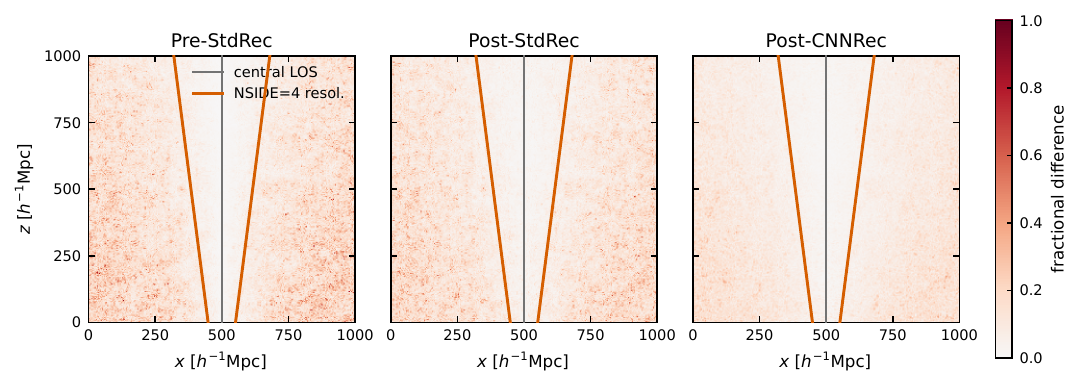}
    \caption{Fractional differences between density fields constructed with a fixed line of sight and with the local lines of sight, averaged over eight subboxes.
    The observer is located at $(x,z)=(500,-400)\,h^{-1}\mathrm{Mpc}$.
    The gray line marks the central line of sight, and the orange lines mark the extent of one \textsc{HEALPix} pixel with \texttt{NSIDE}=4.
    \textbf{Left Panel}: before standard reconstruction.
    \textbf{Middle Panel}: after standard reconstruction.
    \textbf{Right Panel}: after NERV.}
    \label{fig:Appendix_LOS}
\end{figure*}

Figure~\ref{fig:Appendix_LOS} shows the fractional differences between density fields constructed with a fixed line of sight along the $z$-axis and with the local line of sight of each galaxy.
The observer is placed at $(x,z)=(500,-400)\,h^{-1}\mathrm{Mpc}$, so that the subbox covers a distance range similar to that of the first redshift bin.
We compare three stages: the input density field, the field after standard reconstruction, and the field after NERV.
As expected, the difference is smallest near the central line of sight and grows toward the edges, where the local direction deviates more from the $z$-axis.
Within one \texttt{NSIDE}=4 \textsc{HEALPix} pixel, the difference remains modest at all three stages.
The fixed-line-of-sight approximation is therefore accurate enough for the patches used in NERV.

\subsubsection{Redshift-Dependent Number Density}
In a single snapshot of a cubic simulation, the galaxy number density is usually fixed, so the statistical properties of the galaxy distribution are homogeneous along the line of sight. 
This is not the case for realistic galaxy surveys, where the observed number density varies with redshift and may also differ across sky regions due to observation artifacts. 
It is therefore important to test whether NERV remains robust in the presence of such effects.

\begin{figure*}[htbp]
    \centering
    \includegraphics[width=\linewidth]{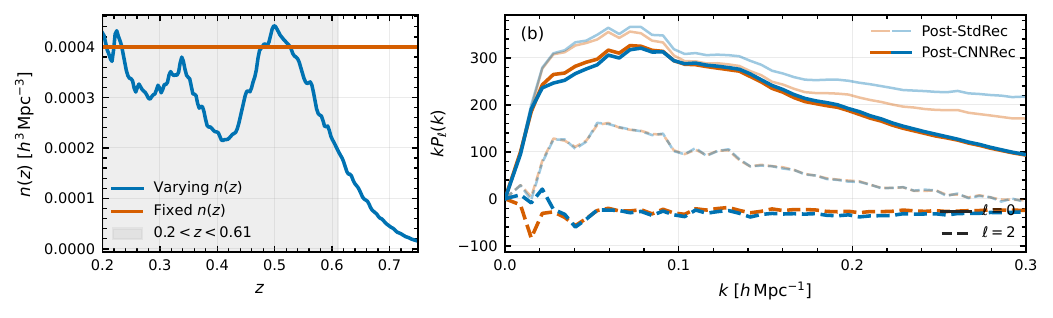}
    \caption{\textbf{Left Panel}: galaxy number density as a function of redshift for the BOSS DR12 catalog (blue) and for the fixed-number-density control sample (orange).
    \textbf{Right Panel}: power spectra after standard reconstruction (transparent) and after NERV (opaque). 
    Solid and dashed lines show the monopole and quadrupole, respectively.
    }
    \label{fig:Appendix_NZ}
\end{figure*}

In Figure~\ref{fig:Appendix_NZ}, we compare the impact of fixed and redshift-dependent number densities on the reconstructed power spectrum. 
The blue curve in the left panel shows the number-density distribution of the BOSS DR12 galaxy catalog. 
We use this distribution within $0.2<z<0.61$ to randomly downsample galaxies in each subbox, with the $z$-axis as the line of sight. 
For comparison, we also generate a control sample with a fixed number density of $n_g=4.0\times10^{-4}\,h^{3}\,\mathrm{Mpc}^{-3}$ (orange curve). 
In both cases, each galaxy is assigned a weight $w=1/n_g$, following the weighting convention used in the real-data analysis.

The right panel of Figure~\ref{fig:Appendix_NZ} shows the power spectra after standard reconstruction and after NERV, with and without the redshift-dependent selection.
Although the imposed number density varies by nearly a factor of two over the redshift range, the power spectra change only at the $\sim10\%$ level.
The change is similar for standard reconstruction and for NERV, so the CNN stage does not amplify the selection effect.
The change is also smooth in $k$, and a smooth change of this size is absorbed by the broadband marginalization in the BAO fit.
We conclude that NERV is robust to realistic redshift-dependent number-density variations.

\subsubsection{Survey Boundary Effects}
\label{sec:boundary}
Survey boundaries are the third observational effect. 
In previous applications to cubic simulations, the reconstruction was performed with periodic boundary conditions, so each subgrid window contains a complete density-field input. 
In realistic surveys, however, boundaries are unavoidable due to the finite footprint and redshift coverage. 
When a reconstruction window approaches these boundaries, part of the input density field may lie outside the observed region, potentially biasing the CNN output. 
It is therefore necessary to quantify the impact of survey boundaries on the reconstructed density field.

\begin{figure}[htbp]
    \centering
    \includegraphics[width=\linewidth]{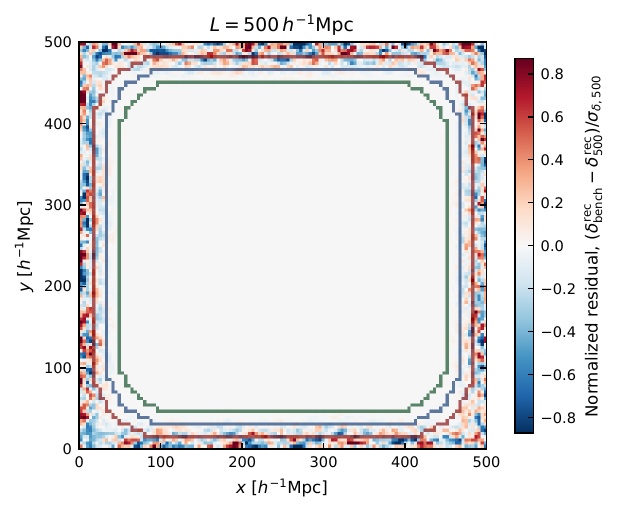}
    \caption{Fractional reconstruction error induced by the survey boundary, computed in a $500\times500\times40\,(h^{-1}\mathrm{Mpc})^3$ slice.
    The red, blue, and green curves mark the boundaries of the effective regions for $\tilde{\gamma}=0.6$, $0.7$, and $0.8$.}
    \label{fig:Appendix_boundary2}
\end{figure}

In Figure~\ref{fig:Appendix_boundary2}, we show the fractional reconstruction error in a local region affected by survey boundaries. 
The local region has a volume of $(500\,h^{-1}\mathrm{Mpc})^3$ and is truncated from a $(1000\,h^{-1}\mathrm{Mpc})^3$ subbox. 
To mimic the unobserved region in a realistic survey, the density field outside this local region is set to zero before applying NERV. 
We then compare the reconstructed field obtained from this truncated region with that obtained from the full subbox.

\begin{figure}[htbp]
    \centering
    \includegraphics[width=\linewidth]{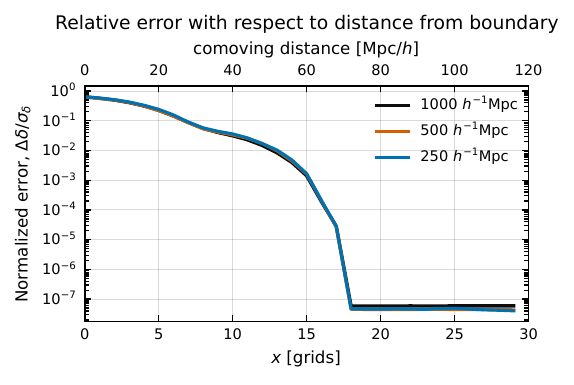}
    \caption{Fractional reconstruction error as a function of distance from the survey boundary.}
    \label{fig:Appendix_boundary1}
\end{figure}

As shown in Figure~\ref{fig:Appendix_boundary2}, the survey boundary does affect the reconstruction, but the induced error is localized near the boundary, mostly within $\sim30\,h^{-1}\mathrm{Mpc}$. 
For a more quantitative comparison, Figure~\ref{fig:Appendix_boundary1} shows the fractional error as a function of the distance from the boundary. 
With the current configuration, the boundary-induced error drops below the percent level at a distance of $50\,h^{-1}\mathrm{Mpc}$, and becomes consistent with numerical precision at $70\,h^{-1}\mathrm{Mpc}$.

To mitigate these effects, we introduce an additional masking step that discards less reliable regions near the survey boundaries. 
For each grid cell $i$, we define an effective volume fraction $\gamma_i$ by centering the CNN subgrid window on that cell and computing the fraction of valid grid cells within the window:
\begin{equation}
\gamma_i = \frac{N_{\mathrm{eff}}^{i}}{N_{\mathrm{tot}}}.
\end{equation}
Here $N_{\mathrm{tot}}$ is the total number of grid cells in the subgrid window, which is $N_{\mathrm{tot}}=50^3$ in this work, and $N_{\mathrm{eff}}^{i}$ is the number of valid grid cells within the window. 
A cell is considered valid if it lies inside the survey volume, i.e., if it contains at least one galaxy or random particle.

We then apply a threshold $\tilde{\gamma}$ to define the effective reconstruction region. 
Grid cells with $\gamma_i < \tilde{\gamma}$ are discarded from power spectrum estimation.
Figure~\ref{fig:Appendix_boundary2} shows the boundaries of the effective regions for thresholds $\tilde{\gamma}=0.6$ (red), $0.7$ (blue), and $0.8$ (green).
The choice $\tilde{\gamma}=0.6$ removes the boundary-dominated grid cells while retaining more of the usable volume than the stricter thresholds.
We therefore adopt $\tilde{\gamma}=0.6$ throughout this work.

\subsection{Effects of Model Discrepancies}
\label{sec:mod_dis}

Simulation-to-data discrepancies can arise in several respects: redshift, galaxy number density, cosmological parameters, galaxy bias, and the galaxy--halo connection.
Some of these have already been tested elsewhere: \citet{Bayer:2026zcr} found hybrid reconstruction robust to an incorrect cosmological model and AP rescaling, and \citet{2025JCAP...09..039P} found that a model trained on halo catalogs performs worse when applied to galaxy catalogs.
Here we test the two remaining sources of discrepancy that can be examined directly in our setup: redshift and galaxy number density.

\subsubsection{Redshift Mismatch}
Figure~\ref{fig:Plot_ModelMisMatch} shows the cross-correlation coefficient between the reconstructed field and the initial conditions, for networks trained at different redshifts and applied to a fixed target redshift.
Models trained at different redshifts perform similarly, so the reconstruction is insensitive to the redshift of the training set.
This supports our choice of training the network at a single fixed redshift and applying it to galaxy samples spanning a finite redshift range.

\subsubsection{Galaxy Number-Density Mismatch}
\label{sec:density_mismatch}
\begin{figure*}
    \centering
    \includegraphics[width=\linewidth]{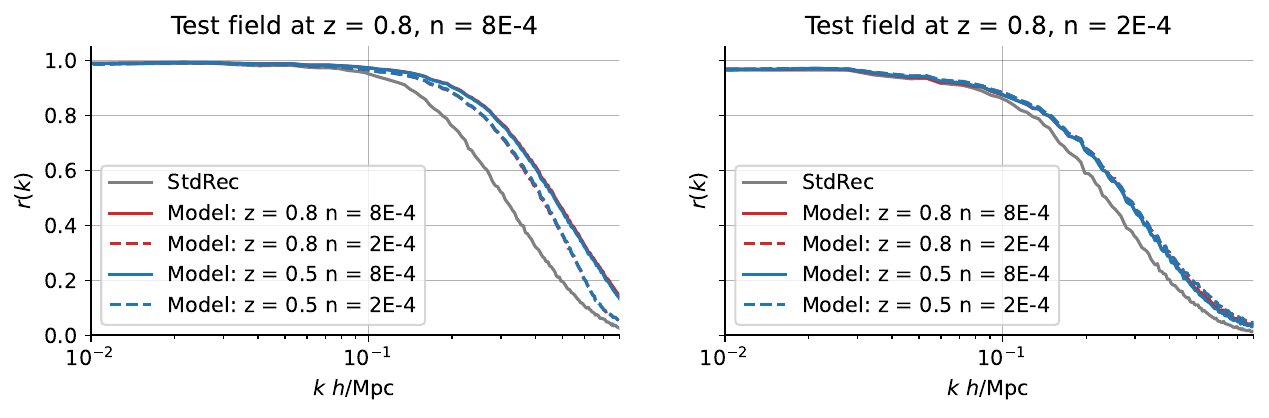}
    \caption{Reconstruction performance for galaxy catalogs with number density $\bar{n}=8\times10^{-4}\,h^3\mathrm{Mpc}^{-3}$ (\textbf{left panel}) and $\bar{n}=2\times10^{-4}\,h^3\mathrm{Mpc}^{-3}$ (\textbf{right panel}).
    Grey curves show standard reconstruction.
    Red and blue curves show NERV models trained at $z=0.8$ and $z=0.5$, respectively.
    Solid curves use models trained on the high-number-density catalog ($\bar{n}=8\times10^{-4}\,h^3\mathrm{Mpc}^{-3}$), and dashed curves use models trained on the low-number-density catalog ($\bar{n}=2\times10^{-4}\,h^3\mathrm{Mpc}^{-3}$).
    }
    \label{fig:Plot_ModelMisMatch}
\end{figure*}

Figure~\ref{fig:Plot_ModelMisMatch} compares the cross-correlation coefficient of reconstructed fields from galaxy catalogs with different number densities, using networks trained on catalogs with different number densities.
On the high-number-density catalog, the model trained at high number density clearly outperforms the model trained at low number density.
On the low-number-density catalog, the two models perform comparably.
We thus note that a model trained at high number density can therefore be applied to lower-density data, while the reverse is less effective.

A likely explanation is that training at high number density lets the network learn more of the small-scale nonlinear structure, which remains useful when applied to sparser data, while a network trained at low number density has less access to this information and struggles to recover it on a denser catalog.
The degeneracy on the low-number-density catalog likely reflects a combination of effects including shot noise limits the available small-scale information, and standard reconstruction itself becomes less effective at low number density, propagating degraded input to the CNN.
Because these effects are coupled, our controlled experiments cannot cleanly separate their individual contributions, and a detailed discussion is beyond the scope of this work.

\section{Results}
\label{sec:results}

In this section, we first evaluate the performance of the NERV using mock catalogs in Section~\ref{sec:validation}.
We then present the reconstruction results for curved-sky mocks and the corresponding BAO fitting results in Section~\ref{sec:results_baofit}.

\subsection{Validation in the Mocks}
\label{sec:validation}

\begin{figure*}
    \centering
    \includegraphics[width=\linewidth]{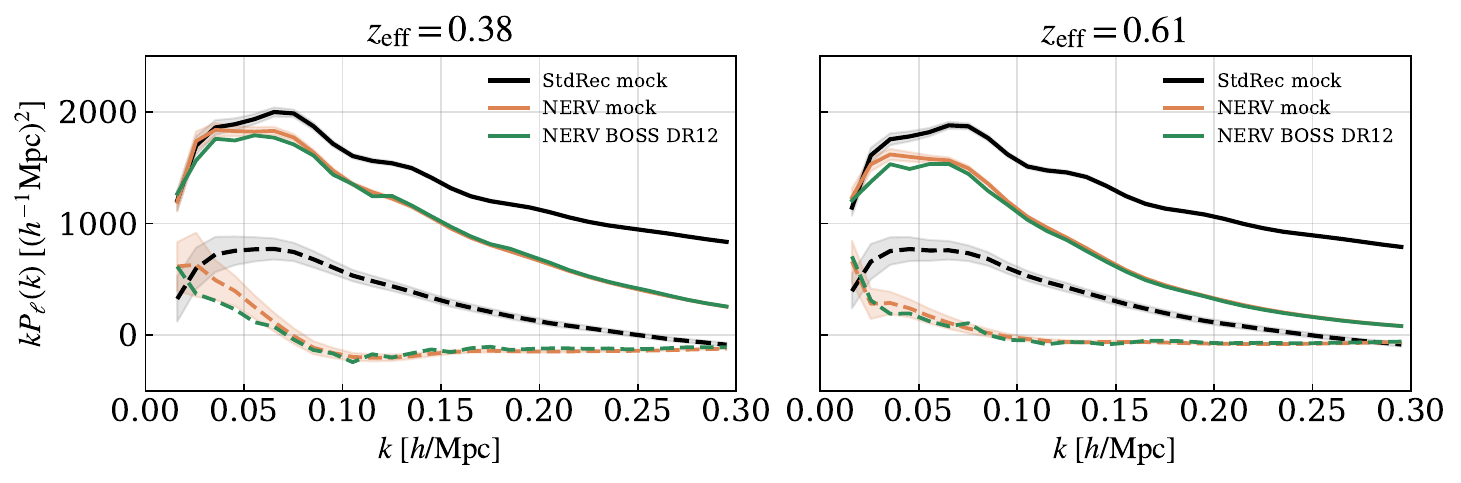}
    \caption{The power spectrum multipoles after reconstruction. 
    The two panels show the results for the two redshift bins. The black curves correspond to standard reconstruction. 
    The orange curves show the NERV results measured from the \textsc{MultiDark-Patchy} mocks, while the green curves show the corresponding measurements from the BOSS DR12 galaxy catalog. 
    The shaded regions indicate the standard deviation of the band powers estimated from 500 mock realizations.
    }
    \label{fig:Fig_Power}
\end{figure*}

Figure~\ref{fig:Fig_Power} shows the reconstructed power spectrum multipoles measured from the mock catalogs.
The NERV preserves the overall shape of the power spectrums.
The quadrupole approaches zero on small scales after NERV, as expected from the isotropic initial density field used as the training target.
A small residual quadrupole remains, which may partially arise from the redshift mismatch between the training sample at $z=0.5$ and the target samples at $z_{\mathrm{eff}}=0.38$ and $0.61$.

Table~\ref{tab:vali_fitting} presents the BAO fitting results obtained from the mean power spectra of 500 mock realizations.
In the first redshift bin, NERV improves the precision of $\alpha_\parallel$ and $\alpha_\perp$ by $\sim8\%$ and $\sim7\%$, respectively, relative to standard reconstruction, while in the second redshift bin the two methods yield statistically indistinguishable precision.
In both bins, the recovered dilation parameters remain consistent with the expected values for the \textsc{MultiDark-Patchy} cosmology, confirming that NERV introduces no bias in the reconstructed BAO signal.
The calibration of the reported uncertainties is verified by a coverage test on the individual mock realizations in Appendix~\ref{sec:coverage}.

\begin{table*}[t]
\centering
\caption{\label{tab:vali_fitting}
Best-fit BAO parameters for the post-standard reconstruction and NERV in two redshift bins with 500 \textsc{MultiDark-Patchy} mocks.}
\begin{ruledtabular}
\begin{tabular}{lcccccc}
& \multicolumn{2}{c}{$0.2 < z < 0.5$}  & \multicolumn{2}{c}{$0.5 < z < 0.75$} \\
& Post-StdRec & NERV  & Post-StdRec & NERV \\
\hline
$\alpha_{\parallel}$ 
& $0.997 \pm 0.0319$ & $0.995 \pm 0.0293$
& $1.001 \pm 0.0287$ & $1.003 \pm 0.0272$ \\

$\alpha_{\perp}$ 
& $0.997 \pm 0.0177$ & $0.997 \pm 0.0164$
& $0.998 \pm 0.0168$ & $0.999 \pm 0.0166$ \\

$\chi^2/\mathrm{d.o.f.}$ 
& $0.75/(58 - 29)$ & $2.78/(58 - 29)$
& $0.73/(58 - 29)$ & $7.44/(58 - 29)$ \\
\end{tabular}
\end{ruledtabular}
\end{table*}

\subsection{BAO fitting result for BOSS DR12}
\label{sec:results_baofit}

\begin{figure*}[htbp]
    \centering
    \includegraphics[width=0.8\linewidth]{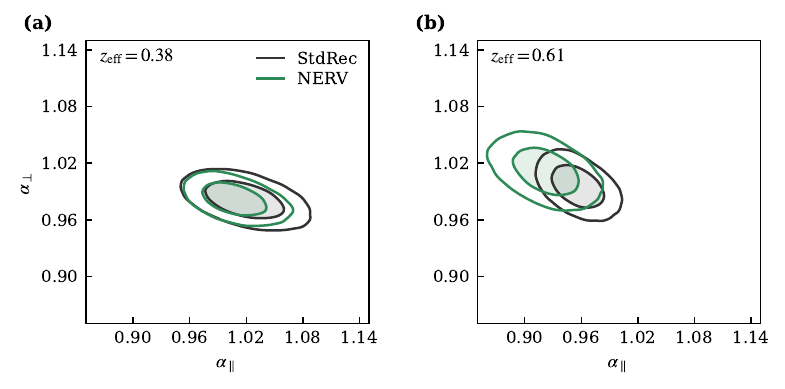}
    \caption{{\bf Constraints on $\alpha_\parallel$ and $\alpha_\perp$ from the reconstructed power spectrum of the BOSS DR12 galaxy sample.}
    The contours indicate the $1\sigma$ and $2\sigma$ confidence regions. }
    \label{fig:Fig2_Contour_data}
\end{figure*}

\begin{table*}[t]
\centering
\caption{\label{tab:bao_results}
Best-fit BAO parameters for the post-standard reconstruction and NERV in two redshift bins with BOSS DR12.}
\begin{ruledtabular}
\begin{tabular}{lcccccc}
& \multicolumn{2}{c}{$0.2 < z < 0.5$}  & \multicolumn{2}{c}{$0.5 < z < 0.75$} \\
& Post-StdRec & NERV  & Post-StdRec & NERV \\
\hline
$\alpha_{\parallel}$ 
& $1.018 \pm 0.0265$ & $1.003 \pm 0.0197$
& $0.955 \pm 0.0170$ & $0.931 \pm 0.0230$ \\

$\alpha_{\perp}$ 
& $0.981 \pm 0.0124$ & $0.982 \pm 0.0109$
& $0.995 \pm 0.0142$ & $1.011 \pm 0.0156$ \\

$\chi^2/\mathrm{d.o.f.}$ 
& $29.46/(58 - 29)$ & $34.55/(58 - 29)$
& $33.85/(58-29)$ & $34.93/(58 - 29)$ \\
\end{tabular}
\end{ruledtabular}
\end{table*}

Figure~\ref{fig:Fig2_Contour_data} and Table~\ref{tab:bao_results} present a comparison of the BAO dilation-parameter constraints obtained with standard reconstruction and NERV for the BOSS DR12 galaxy sample.
In the first redshift bin, standard reconstruction yields
$\alpha_\parallel = 1.018 \pm 0.0265$ and $\alpha_\perp = 0.981 \pm 0.0124$,
while NERV gives
$\alpha_\parallel = 1.003 \pm 0.0197$ and $\alpha_\perp = 0.982 \pm 0.0109$,
improving the precision of $\alpha_\parallel$ and $\alpha_\perp$ by $\sim25\%$ and $\sim12\%$, respectively. 
The magnitude of this improvement is generally consistent with that found in previous studies of CNN-based reconstruction \citep[][]{2025JCAP...09..039P}.
These results indicate that the additional information recovered by the CNN can translate into tighter constraints on the BAO dilation parameters in observational data, though the size of this gain depends on the number density of the sample.

In the second redshift bin, standard reconstruction yields
$\alpha_\parallel = 0.955 \pm 0.0170$ and $\alpha_\perp = 0.995 \pm 0.0142$,
while NERV gives
$\alpha_\parallel = 0.931 \pm 0.0230$ and $\alpha_\perp = 1.011 \pm 0.0156$;
here NERV yields no improvement, and its uncertainties are in fact larger than those of standard reconstruction.
The absence of improvement in the second bin is also seen in the mock validation. 

We attribute the lack of improvement in the second redshift bin to two related factors.
First, this bin has a lower number density and thus higher shot noise, which limits the small-scale nonlinear information available to the network.
Second, the network for this bin is trained on catalogs downsampled to $\bar{n}=2.0\times10^{-4} \ h^3\mathrm{Mpc}^{-3}$ to match the mean number density of the galaxy sample, whereas parts of the BOSS DR12 sample's number density in this bin is higher.
As shown in Section~\ref{sec:density_mismatch}, a network trained at low number density has limited access to small-scale information and does not fully exploit denser input, so this mismatch further limits the achievable gain.
We leave improving the reconstruction performance in this redshift bin to future work.

\section{Discussion}
\label{sec:discussion}

The key result of this work is that the NERV improves BAO constraints in a real survey.
When applied to the BOSS DR12 galaxy sample, the method improves the precision of the BAO distance measurements by up to $\sim25\%$ in the lower-redshift bin, while the higher-redshift bin shows no improvement relative to standard reconstruction.
This gain suggests that NERV recovers additional nonlinear information that is not fully captured by the traditional method.

Previous machine-learning-based reconstruction methods, 
\citep{Mao:2020vdp, Shallue:2022mhf, Chen:2023uup, 2025JCAP...09..039P, Bayer:2026zcr}, have been developed and validated exclusively in periodic simulation boxes.
The results presented here show that the same method can be deployed on an irregularly shaped, curved-sky galaxy survey with a redshift-dependent selection function, provided the geometric and observational effects are handled explicitly, as demonstrated in Section~\ref{sec:robustness}.
This tessellation-and-masking approach is independent of the specific network architecture and could equally be applied to the methods above, offering a practical route for the broader machine-learning-reconstruction community to move from simulations to real survey data.

While NERV yields a clear improvement in the first redshift bin, the lack of improvement at higher redshift highlights a limitation in the current model's ability to generalize beyond its training distribution.
At $z_{\mathrm{eff}}=0.38$ the number density of the sample is nearly constant, so the data lie within the regime covered by the training set and the network performs as intended.
At higher redshift, however, the reconstruction performance degrades, suggesting that the data depart from the regime on which the model was trained. 
Given the nonlinear nature of the network, it is difficult to isolate the dominant source of this mismatch, although the mismatch between the number-density distribution in the training set and that of the observed sample is a likely contributor.
Addressing this limitation will require training samples that more closely reproduce the distribution of the observed data, together with controlled tests designed to identify which aspects of the mismatch are primarily responsible for the degradation.

We note that the best-fit $\chi^2$ values obtained after NERV are somewhat larger than those from standard reconstruction in both redshift bins, consistent with the same pattern seen in the mock validation.
A possible explanation is that in the present analysis, we directly adopt the same power-spectrum template used for the standard reconstruction without re-calibration.
This increase in $\chi^2$ therefore likely reflects a modest modeling mismatch rather than a degradation in the BAO information.
A detailed template design based on mocks may further improve the fit and provide a more accurate description of the reconstructed power spectrum.

Finally, the number densities of DESI luminous red galaxies and emission-line galaxies are generally higher and more uniform across redshift than those of BOSS DR12, including over the redshift range corresponding to our higher-redshift bin. 
We therefore expect NERV-like reconstruction to provide a robust improvement in BAO precision across the DESI redshift range as reported by \citet[][]{Bayer:2026zcr}. 
Extending this approach to DESI and future spectroscopic surveys could further sharpen cosmological distance measurements and improve sensitivity to fundamental physics, including the nature and evolution of dark energy.

\begin{acknowledgments}

We thank Meng Zhou and Adrian Bayer for helpful comments that significantly improved the manuscript. 
S.-H. Z. additionally thanks Adrian Bayer and Uro\v{s} Seljak for helpful discussions.
H.-M. Z. acknowledges support from the National Key R\&D Program of China (2025YFA1614103, 2023YFA1607800, and 2023YFA1607803) and the National Natural Science Foundation of China (NSFC; Grant No. 12622301).
U.-L. P. receives support from the Canadian Institute for Advanced Research Fund (CIFAR Fund), the Natural Sciences and Engineering Research Council of Canada (NSERC) [funding reference numbers RGPIN-2019-06770, ALLRP 586559-23], and the Ontario Research Fund-Research Excellence (ORF-RE), the Academia Sinica [Grand Challenge Seed Program AS-GCS-114-M02 and Investigator Award AS-IV-115-M04], and the National Science and Technology Council (NSTC) of Taiwan [grant no. 115-2123-M-001-009].
Research at the Perimeter Institute is supported in part by the Government of Canada through the Department of Innovation, Science and Economic Development and by the Province of Ontario through the Ministry of Colleges and Universities.
Computations were performed on the Trillium supercomputer at the SciNet HPC Consortium. SciNet is funded by Innovation, Science and Economic Development Canada; the Digital Research Alliance of Canada; the Ontario Research Fund: Research Excellence; and the University of Toronto.
Claude and Codex were used to assist with language editing during the preparation of this manuscript.

\end{acknowledgments}

\appendix

\section{Coverage test}
\label{sec:coverage}

\begin{figure*}
    \centering
    \includegraphics[width=0.6\linewidth]{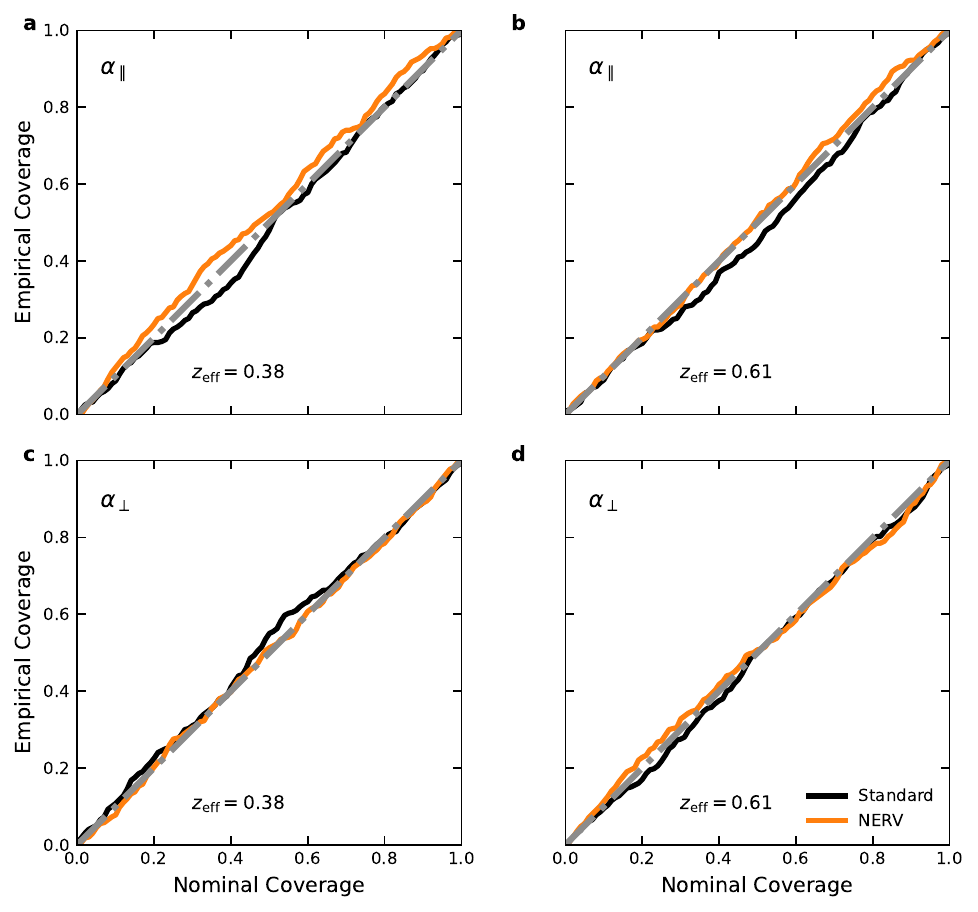}
    \caption{Coverage test of the BAO dilation-parameter uncertainties using 500 individual fits on \textsc{MultiDark-Patchy} mock realizations.
    Each panel shows the empirical coverage against the nominal coverage, for $\alpha_\parallel$ (\textbf{a}, \textbf{b}) and $\alpha_\perp$ (\textbf{c}, \textbf{d}) at $z_{\mathrm{eff}}=0.38$ (left) and $z_{\mathrm{eff}}=0.61$ (right).
    Black and orange curves show standard reconstruction and NERV respectively.}
    \label{fig:coverage}
\end{figure*}

Since an apparent reduction in parameter uncertainties is only meaningful if correctly calibrated, as emphasized in previous validation studies~\cite{Bayer:2026zcr}, we perform a coverage test using 500 \textsc{MultiDark-Patchy} mock realizations fitted individually. 
For each nominal confidence level, we compute the empirical coverage as the fraction of realizations whose credible interval contains the expected value of the corresponding dilation parameter.
Figure~\ref{fig:coverage} shows the result.
Overall, the empirical coverage follows the nominal confidence level well for both methods.
In the low-redshift bin, the constraints are well calibrated, although the NERV $\alpha_\parallel$ uncertainty is slightly conservative.
Given the finite number of mock realizations, the statistical significance of these small deviations remains to be quantified.

\bibliography{sample701}{}
\bibliographystyle{aasjournalv7}

\end{document}